\documentclass[12pt,preprint]{aastex}
\usepackage{bm}

\shorttitle{The missing link spectrum }
\shortauthors{Teraki et al.}

\begin{document}

\title{A NOVEL EMISSION SPECTRUM FROM A RELATIVISTIC ELECTRON MOVING IN A RANDOM MAGNETIC FIELD}

\author{Yuto Teraki\altaffilmark{1} and Fumio Takahara\altaffilmark{1}}
\affil{Department of Earth and Space Science, Graduate School of Science, Osaka University, 1-1 Machikaneyama-cho, Toyonaka, Osaka 560-0043, Japan}
\email{teraki@vega.ess.sci.osaka-u.ac.jp}

\begin{abstract}
We calculate numerically the radiation spectrum from relativistic electrons moving in small scale 
turbulent magnetic fields expected in high energy astrophysical sources.
Such radiation spectrum is characterized by the strength parameter
$a = \lambda_{\mathrm{B}} e|B|/mc^2$, where $\lambda_{\mathrm{B}}$ is the length scale of the turbulent field.
When $a$ is much larger than the Lorentz factor of a radiating electron $\gamma$, synchrotron radiation is realized,
while $a\ll 1$ corresponds to the so-called jitter radiation regime.
Because for $1<a<\gamma$ we cannot use either approximations, 
we should have recourse to the Lienard-Wiechert potential to evaluate the radiation spectrum, which is 
performed in this paper.
We generate random magnetic fields assuming Kolmogorov turbulence,
 inject monoenergetic electrons, solve the equation of motion, and calculate the radiation spectrum. 
We perform numerical calculations for several values of $a$ with $\gamma = 10$. 
We obtain various types of spectra ranging between jitter radiation and synchrotron radiation.
For $a\sim 7$, the spectrum turns out to take a novel shape which has not been noticed up to now.
It is like a synchrotron spectrum in the middle energy region, but in the low frequency 
region it is a broken power law and 
in the high frequency region an extra power law component appears beyond the synchrotron cutoff.
We give a physical explanation of these features.   
\end{abstract}


\section{Introduction}
A major part of non-thermal emission from high energy astrophysical objects is almost always characterized by
the radiation from relativistic electrons moving in magnetic fields.
Usually it is interpreted in terms of the synchrotron radiation.
However, synchrotron approximation is not always valid, in particular when the magnetic fields are highly turbulent.
Electrons suffer from random accelerations and do not trace a helical trajectory.
In general, the radiation spectrum is characterized by the strength parameter
\begin{equation}
a= \lambda_{\mathrm{B}} \frac{e|B|}{mc^2},
\end{equation}
where $\lambda_{\mathrm{B}}$ is the typical scale of turbulent fields, $|B|$ is the mean value of the turbulent magnetic fields, 
$e$ is the elementary charge, 
$m$ is the mass of electron and $c$ is the speed of light (Reville \& Kirk 2010).
When $a\gg \gamma$, where $\gamma$ is the Lorentz factor of radiating electron, 
 the scale of turbulent fields is much larger than the Larmor radius $r_g\equiv \gamma m c^2/e|B|$, 
and electrons move in an approximately uniform field, so that the synchrotron approximation is valid.
In contrast, when $a\ll1$, $\lambda_{\mathrm{B}}$ is much smaller than the scale $r_g/\gamma = \lambda_{\mathrm{B}}/a$ which 
corresponds to the emission of the characteristic synchrotron frequency. 
In this regime, electrons move approximately straightly, and jitter approximation 
or the weak random field regime of diffusive synchrotron radiation (DSR) can be applied (Medvedev 2010, Fleishman and Urtiev 2010).
For $1\lesssim a \lesssim \gamma$, no simple approximation of the radiation spectrum has been known.

The standard model of Gamma Ray Bursts (GRB) is based on the synchrotron radiation 
from accelerated electrons at the internal shocks.
The observational spectra of prompt emission of GRB can be well fitted by a broken power law spectrum which is called the Band function.
Around a third of GRBs show a spectrum in the low energy side harder than the synchrotron theory predicts.
To explain this, other radiation mechanisms are needed.
Medvedev examined relativistic collisionless shocks in relevance to internal shocks of GRB, 
and noticed the generation of small scale turbulent magnetic fields near the shock front (Medvedev \& Loeb 1999).
Then he calculated analytically radiation spectrum from electrons moving in small scale turbulent magnetic fields,
to make a harder spectrum than the synchrotron radiation (Medvedev 2000).
However, he assumed that the strength parameter $a$ is much smaller than $1$ 
and that turbulent field is of one-dimensional structure, 
which may be over simplified in general (Fleishman 2006).
Medvedev also calculated 3-dimensional structure assuming that
the turbulent field is highly anisotropic (Medvedev 2006).
He conclude that the harder spectrum is achieved in "head on" case,
and that in "edge on" case, the spectrum is softer than synchrotron radiation. 
The spectral index depends on the angle $\theta$ between the particle velocity and shock normal with 
hard spectrum obtained when $\theta \lesssim 10^{\circ}$ (Medvedev 2009).
 
Recently several particle-in-cell (PIC) simulations 
of relativistic collisionless shocks have been performed to study the nature of turbulent magnetic fields
which are generated near the shock front (e.g., Frederiksen et al. 2004; Kato 2005; Chang et al. 2008; 
Haugbolle 2010).
The characteristic scale of the magnetic fields is the order of skin depth 
as predicted by the analysis of Weibel instability.
Then, the wavelength of turbulent magnetic field $\lambda_{\mathrm{B}}$ is described by using a coefficient $\kappa$ as
\begin{equation}
\lambda_{\mathrm{B}} = \kappa  \frac{c}{\omega_{\mathrm{pe}} \Gamma_{\mathrm{int}}},
\end{equation}
where $\omega_{\mathrm{pe}}$ is the plasma frequency, and $\Gamma_{\mathrm{int}}$ is the relative Lorentz factor between colliding shells.
The energy conversion fraction into the magnetic fields
\begin{equation}
\varepsilon_{\mathrm{B}} = \frac{B^2/8\pi}{\Gamma_{\mathrm{int}} n m_{\mathrm{p}} c^2}
\end{equation}
is $10^{-3}-0.1$, where $B^2/8\pi$ is energy density of magnetic fields, and
 $\Gamma_{\mathrm{int}} n m_{\mathrm{p}} c^2$ is the kinetic energy density of the shell.
The Lorentz factor of electrons is similar to $\Gamma_{\mathrm{int}}$,
and that $\kappa$ is typically $10$ from the result of PIC simulations, 
the strength parameter $a$ can be estimated as
\begin{equation}
a = \sqrt{2}\kappa{\varepsilon_{\mathrm{B}}}^{1/2}\sqrt{\frac{m_{\mathrm{p}}}{{m_{\mathrm{e}}}}}
\sim O(1-10^2).
\end{equation}
Thus, the assumption $a\ll1$ on which jitter radiation and DSR weak random field regime
are based is not necessarily valid when we consider the radiation from the internal shock region of GRB.

Fleishman \& Urtiev (2010) calculated the radiation spectrum for $a>1$ using a statistical method, 
but their treatment of the "small scale component" is somewhat arbitrary.
They introduced the critical wavelength $\lambda_{\mathrm{crit}}$ and called
 components with $\lambda \leq \lambda_{\mathrm{crit}}$ the "small scale component", 
where $\lambda_{\mathrm{crit}}$ obeys 
the inequality
\begin{equation}
r_{\mathrm{g}} \ll \lambda_{\mathrm{crit}} \ll r_{\mathrm{g}}/\gamma
\label{crit}
\end{equation}
(Toptygin \& Fleishman 1987). 
The inequality can be transformed to $1\ll \lambda_{\mathrm{crit}}e|B|/mc^2 \ll \gamma$,
so that the division through $\lambda_{\mathrm{crit}}$ may be
 ambiguous when we calculate the radiation spectrum for $1<a<\gamma$. 

The synthetic spectra from PIC simulations were calculated recently 
(e.g., Hededal 2005; Sironi \& Spitkovsky 2009; 
Frederiksen 2010; Reville \& Kirk 2010; Nishikawa et al 2011). 
Althogh their magnetic fields are realistic and self-consistent,
it is inevitable that the fields are described by discrete cells in PIC simulation.
Reville \& Kirk (2010) developed an alternative method of calculation of radiation spectra that uses 
the concept of photon formation length, which costs much shorter time than the first principle method
utilizing the Lienard-Wiechert potential.

In this paper, we rather use the first principle method to obtain the spectrum as exact as possible.
We adopt the field description method developed    
by Giacalone \& Jokipii (1999) and used by Reville \& Kirk (2010).
We assume isotropic turbulent magnetic fields which have broader 
power spectra $k_{\mathrm{max}}=100 \times k_{\mathrm{min}}$
and calculate the radiation spectra in the regime of $1<a<\gamma$.
In \S 2 we describe calculation method and numerical results. 
In \S 3 we give a physical interpretation.

\section{Method of calculation}
Because we focus our attention on calculating radiation spectrum, we assume the static field with required properties of $a$,
and neglect the back reaction of radiating electrons to the magnetic field.
We solve the trajectory of electron accurately in each time step and calculate the radiation spectrum.
\subsection{Setting}
The isotropic turbulent field is generated by using the discrete 
Fourier transform description as developed in Giacalone \& Jokipii (1999).
It is described as a superposition of \textit{N} Fourier modes, 
each with a random phase, direction and polarization
\begin{equation} 
\bm{B}(\bm{x}) = \sum_{n=1}^N A_n \exp\bigl\{i(\bm{k}_n \cdot \bm{x} + \beta_n)\bigr\}\hat{\xi}_n. 
\end{equation}
Here, $A_n,\beta_n,\bm{k}_n \: \mathrm{and} \: \hat{\xi}_n$ are the amplitude, phase, 
wave vector and polarization vector for the \textit{n}-th mode, respectively.
The polarization vector is determined by a single angle $0<\psi_n<2\pi$
\begin{equation} 
\hat{\xi}_n = \cos{\psi_n}\bm{e_x^{\prime}} + i \sin{\psi_n}\bm{e_y^{\prime}},  
\end{equation}
where $\bm{e_x^\prime}$ and $\bm{e_x^\prime}$ are unit vectors, orthogonal to $\bm{e_z^{\prime}} = \bm{k_n}/k_n$.
The amplitude of each mode is
\begin{equation}
A_n^2 = \sigma^2 G_n \Biggl[\sum_{n=1}^N G_n \Biggr]^{-1}, 
\end{equation}
where the variance $\sigma$ represents the amplitude of the turbulent field.
We use the following form for the power spectrum
\begin{equation}
G_n = \frac{4 \pi k_n^2 \Delta k_n}{1 + (k_n L_{\mathrm{c}})^{\alpha}}, 
\end{equation}
where  $L_c$ is the correlation length of the field. Here, $\Delta k_n$ is chosen such that there is an equal spacing in logarithmic
$k-$space, over the finite interval $k_{\mathrm{min}} \leqq k \leqq k_{\rm{max}}$, where $k_{\rm{max}} = 100 \times k_{\rm{min}}$ and $N = 100$.
where $k_{\rm{min}} = 2\pi/L_{\mathrm{c}}$ and $\alpha = 11/3$.
We have no reliable constraint for value of $\alpha$ from GRB observation, 
so we adopt the Kolmogorov turbulence $B^2(k) \propto k^{-5/3}$ tentatively,
where the power spectrum has a peak at $k_{\mathrm{min}}$.
Then we define the strength parameter using $\sigma$ and $k_{\rm{min}}$ as
\begin{equation}
a \equiv \frac{2\pi e \sigma}{mc^2k_{\mathrm{min}}}. 
\end{equation}
We inject isotropically 32 monoenergetic electrons with $\gamma=10$ in the prescribed magnetic fields,
and solve the equation of motion 
\begin {equation}
\gamma m_{\mathrm{e}} \dot{\bm{v}} = -e \bm{\beta}\times\bm{B}
\end{equation}
using 2nd order Runge-Kutta method. 
We pursue the orbit of electrons up to the time $300 \times T_{\mathrm{g}}$ , 
where $T_{\mathrm{g}}$ is the gyro time $T_{\mathrm{g}} = 2\pi \gamma m c / e \sigma$.
We calculate radiation spectrum using acceleration $\dot{\bm{v}} = \dot{\bm{\beta}}c$.
The energy $dW$ emitted per unit solid angle $d\Omega$ (around the direction $\bm{n}$) and per unit frequency $d\omega$ to the direction $\bm{n}$ is computed as
\begin{equation} 
\frac{dW}{d\omega d\Omega} = \frac{e^2}{4 \pi c^2} 
\Bigl| \int^{\infty}_{-\infty} \:dt^{\prime} \frac{ \bm{n} \times \bigl[ (\bm{n} - \bm{\beta}) \times \dot{\bm{\beta}} \bigr] } 
 {(1 - \bm{\beta} \cdot \bm{n} )^2 }\exp\bigl\{{i\omega ( t^{\prime} - \frac{\bm{n} \cdot \bm{r}(t^{\prime})}{c})}\bigr\} \Bigr|^2,
\end{equation}
where $\bm{r}(t^{\prime})$ is the electron trajectory, $t^{\prime}$ is retarded time (Jackson 1999).

\subsection{Results}
\begin{figure}
\epsscale{.80}
\includegraphics[angle=270]{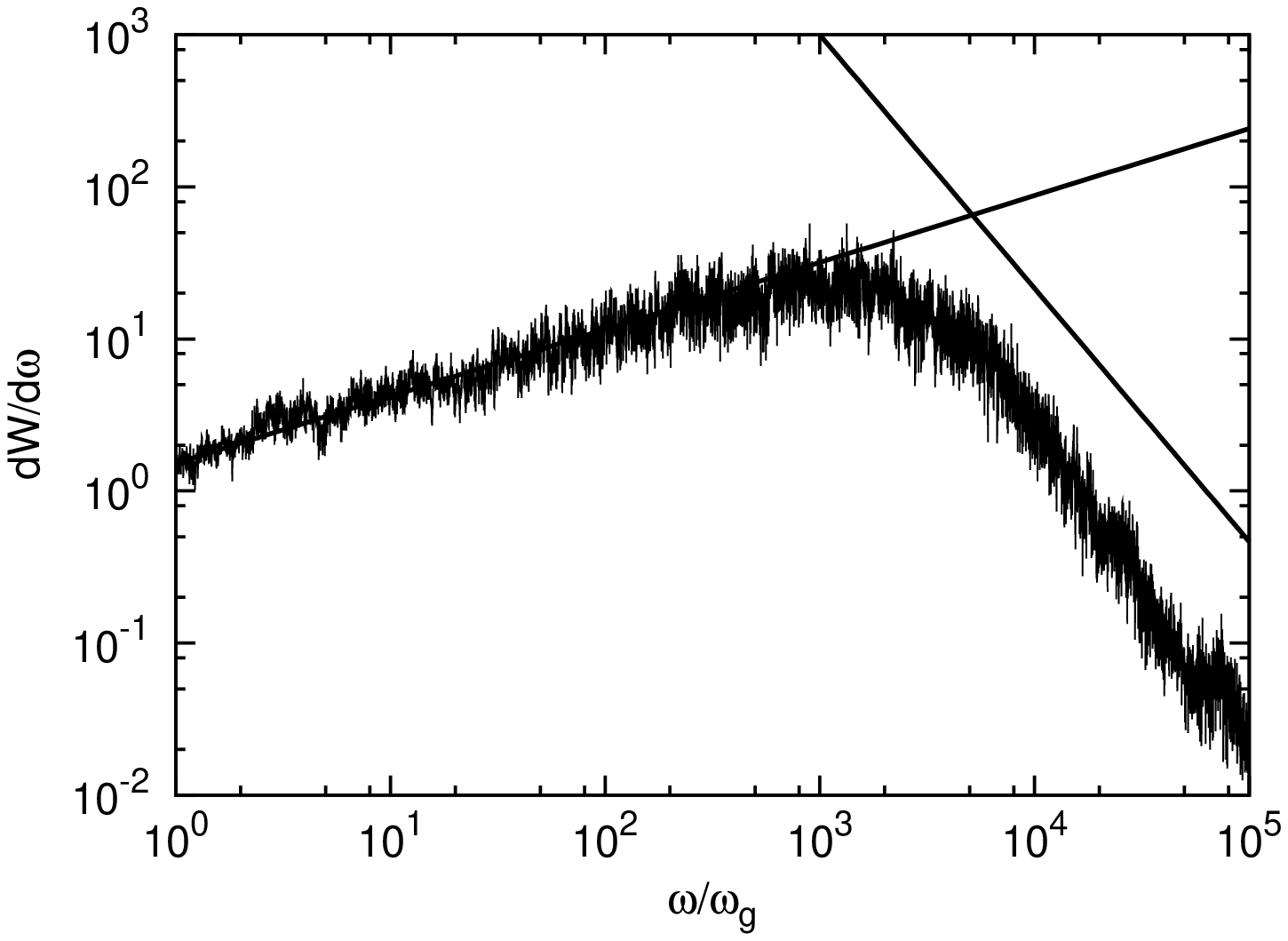}
\caption{Radiation spectrum for $a=3, \: \alpha = 11/3 \: \mathrm{and}\: \gamma=10$.
The straight line in the low frequency region shows a power law spectrum with an index 0.44.
The straight line in the high frequency region is $dW/d\omega \propto \omega^{-5/3}$ for reference.
Power law index of the low frequency spectrum is harder than the synchrotron theory predicts.}
\label{a3}
\end{figure}

First, we show the radiation spectrum for $a = 3$ in Figure \ref{a3}.
The frequency is normalized by the fundamental frequency $\omega_{\mathrm{g}} = e\sigma/(\gamma mc)$, and 
the magnitude is arbitrarily scaled.
The jagged line is the calculated spectrum, while the straight line drawn in the low frequency region 
is a line fitted to a power law spectrum. 
The fitting is made in the range of $1-350\omega_{\mathrm{g}}$ and the spectral index turns out to be $0.44$. 
The straight line drawn in the high frequency region shows a spectrum of $ \propto \omega^{-5/3}$ 
expected for diffusive synchrotron radiation for reference (Toptygin \& Fleishman 1987).
The spectrum is well described by a broken power law, and the spectral index of the low energy side is 
harder than synchrotron theory predicts.
The peak frequency of this spectrum is located at around $ 10^3 \omega_{\mathrm{g}}$.
This frequency corresponds to approximately the typical frequency of synchrotron radiation 
$\omega_{\mathrm{syn}} =3\gamma^2 e\sigma/2mc \sim 10^3 \omega_{\mathrm{g}}$, for $\gamma =10$.

\begin{figure}
\epsscale{.80}
\includegraphics[angle=270]{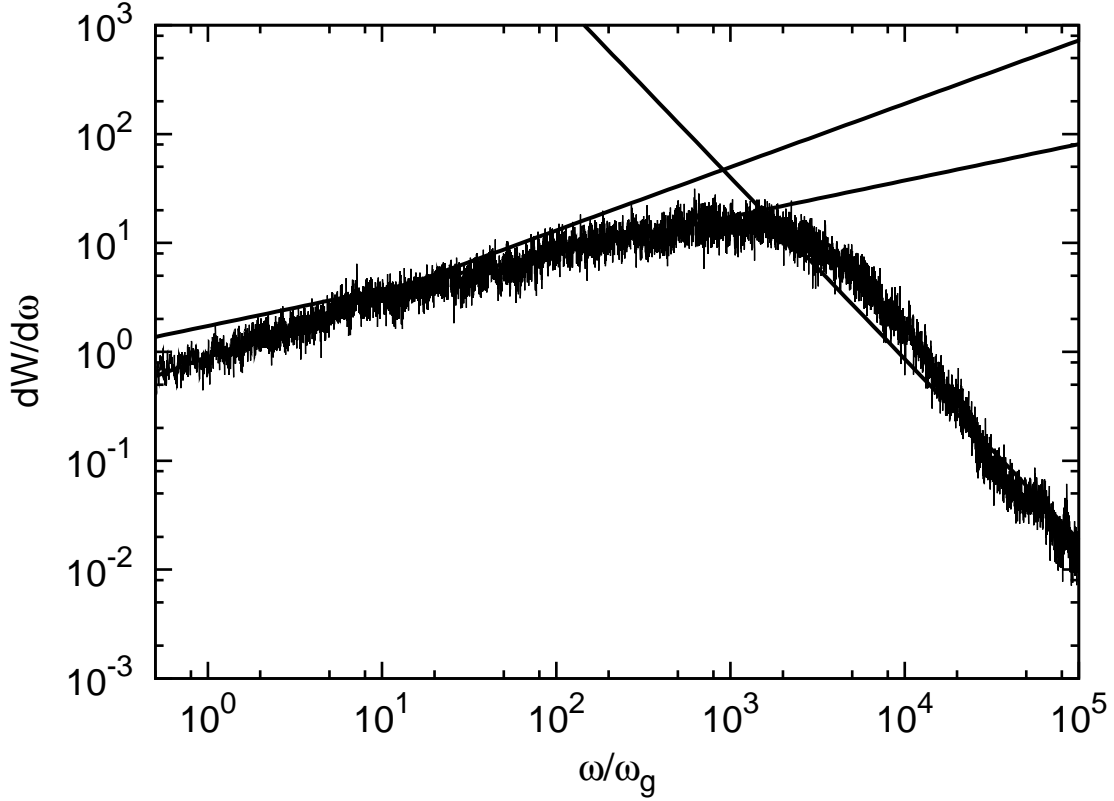}
\caption{Radiation spectrum for $a=5,\alpha = 11/3 \: \mathrm{and} \: \gamma = 10$. 
Two straight lines drawn in low frequency region are fitted one to a 
power law spectrum in the range of $0.5-10\omega_{\mathrm{g}}$ and 
 $10-10^3\omega_{\mathrm{g}}$, respectively. 
The power law index is 0.58 and 0.33.
The latter corresponds to the synchrotron radiation.
The straight line in the high frequency region is $dW/d\omega \propto \omega^{-5/3}$, and
we see a broad hump in the peak region, which is identified with the synchrotron spectrum with an exponential cutoff.}
\label{a5}
\end{figure}

Figure \ref{a5} shows the spectrum for $a = 5$.
The spectral shape changes from that of $a=3$ in both sides of the peak.
The spectrum of the low frequency side becomes a broken power law with a break around $10\omega_{\mathrm{g}}$,
 above which the spectrum is fitted by a power law with an index of $0.33$, 
as expected for synchrotron radiation, while below the break the index is $0.58$.
The high frequency side above the peak indicates an excess above a power law spectrum $dW/d\omega d\Omega \propto \omega^{-5/3}$.
It looks like an exponential cutoff.
The whole spectrum is described by a superposition of a synchrotron spectrum and a DSR broken power law spectrum.
This spectral shape is totally novel and is different from the one by described Fleishman (2010).
He reported that the spectrum is described by broken power law in the same range of $a$ as this work
 ($1<a<\gamma,\: a\sim 10^2 \:\mathrm{and}\: \gamma\sim 10^3$) (Fleishman 2010).

To confirm our inference we calculate the case of $a=7$, for three different values of $\alpha$, i.e.,
$\alpha = 14/3,\: 11/3 \: \mathrm{and}\: 8/3$ and the results are shown in Figure \ref{a7}.
The curved black line is a theoretical curve of synchrotron radiation, and three straight black lines are
$dW/d\omega d\Omega \propto \omega^{-2/3}, \omega^{-5/3} \: \mathrm{and} \: \omega^{-8/3}$ expected for DSR theory for reference. 
The green line corresponding to $\alpha = 14/3$ reveals a clear exponential cutoff, and reveals DSR component in only the highest frequency region.
The power law index of this component in the highest frequency 
region coincides with the expected value $-\alpha + 2 = -8/3$.
The red and blue lines correspond to $\alpha = 11/3 \: \mathrm{and}\: 8/3$, respectively. 
They indicate a common feature to the green one. 

\section{Interpretation}
We give a physical explanation of the spectra obtained in the previous section.
At first, we interpret the broken power law spectrum for $a=3$ by using DSR theory.
Next, we consider physical interpretation of the complex shape of radiation 
spectrum for $a=5$ and $a=7$.
Finally, we compare our work with previous studies.

To begin with, we review the spectral feature of DSR based on the  non-perturbative approach for $a<1$ (Fleishman 2006).
The typical spectrum takes a following form:
$dW/d\omega d\Omega \propto \omega^{1/2}$ in the low frequency region, $\propto \omega^{0}$ in the middle frequency region,
and $\propto \omega^{-\alpha + 2}$ in the high frequency region.
The low frequency part and the middle frequency part are separated 
at $\omega_{\mathrm{LM}} \sim a \omega_{\mathrm{syn}}$.
This spectral break corresponds to the break of the straight orbit 
approximation due to the effect of multiple scattering (Fleishman 2006).
On the other hand, the middle frequency part and the high frequency part are separated at 
the typical frequency of jitter radiation $\omega_{\mathrm{MH}} = \omega_{\mathrm{jit}}$ which is estimated by using the method of virtual quanta as
$\omega_{\mathrm{jit}} \sim \gamma^2 k_{\mathrm{min}}c \sim a^{-1} \omega_{\mathrm{syn}}$ 
(Medvedev 2000, Rybicki \& Lightman 1979).
Then, for $a \sim 1$, $\omega_{\mathrm{LM}} \sim \omega_{\mathrm{MH}}$ is achieved and the middle region may vanish.
The spectrum for $a\sim1$ becomes a broken power law with only one break, which is located at roughly the synchrotron frequency $\omega_{\mathrm{syn}}$.
The power law index of low frequency side is $\sim 0.5$ (which is harder than synchrotron radiation), and that of high frequency side is $-\alpha +2= -5/3$.
Thus, the spectral feature for $a=3$ can be explained by an extrapolation of DSR non-perturbative approach for $a<1$,
if we consider that the middle frequency region is not conspicuous.
Although our spectral index $0.44$ slightly differs from $0.5$ for DSR,
 this index is still harder than the synchrotron theory.
The situation $a\sim1$ can be achieved at the internal shock region of GRB, 
so that this may be responsible for harder spectral index than synchrotron observed for some GRBs.

\begin{figure}
\epsscale{.80}
\plotone{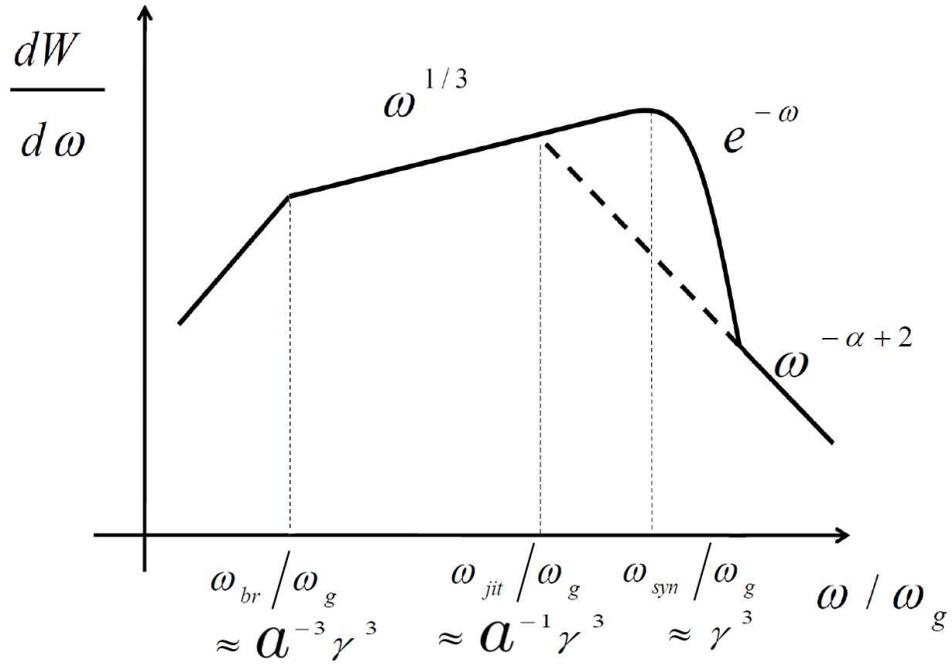}
\caption{Conceptual diagram of the radiation spectrum for $5\leq a<\gamma$.}
\label{spece}
\end{figure}

\begin{figure}
\epsscale{.80}
\includegraphics[scale=0.7, angle=90]{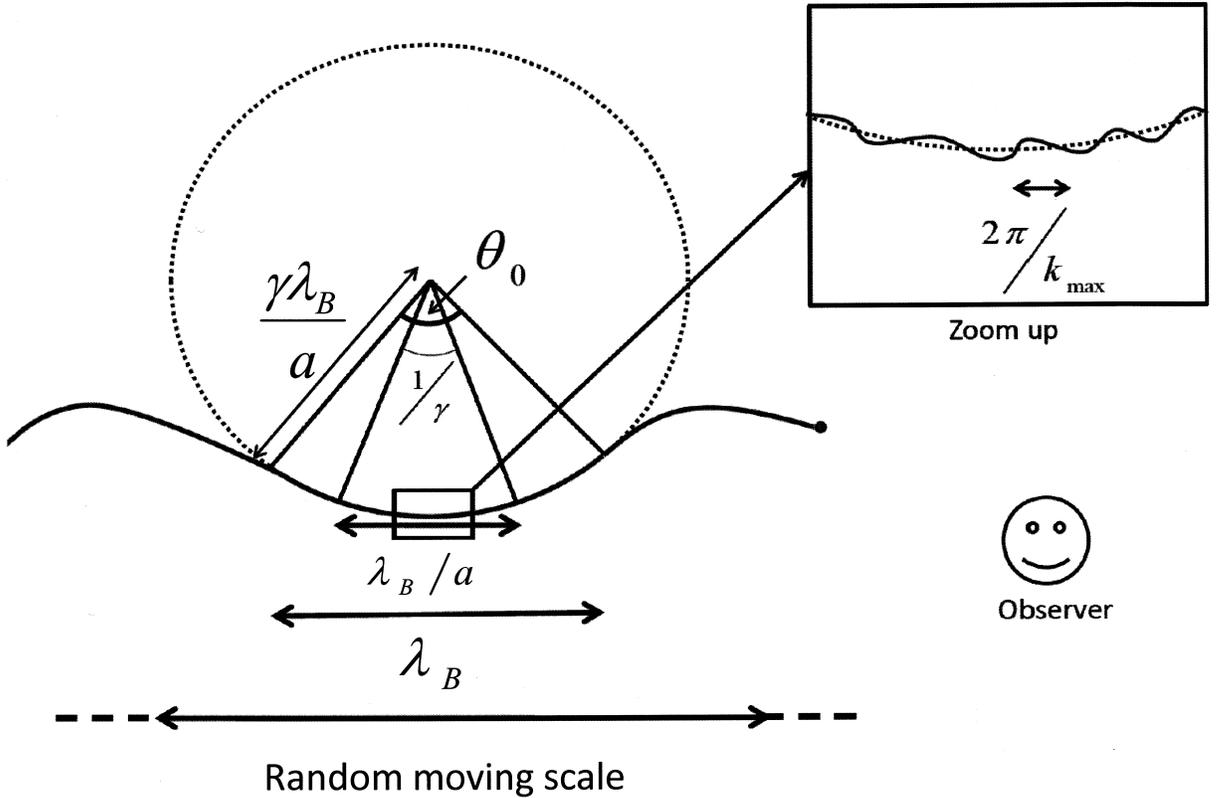}
\caption{Cartoon showing the electron trajectory for $1<a<\gamma$.
The radius of the guiding circle is the Larmor radius $r_{\mathrm{g}}= \gamma\lambda_{\mathrm{B}}/a$.
Low frequency photons are emitted from the motion on relatively scales larger than 
$\lambda_{\mathrm{B}}$, while middle frequency ones from that on the intermediate scale 
between $\lambda_{\mathrm{B}}/a$ and $\lambda_{\mathrm{B}}$ are basically synchrotron radiation.
The spectral break at $a^{-3}\gamma^3$ in Figure \ref{spece} corresponds to the 
break of synchrotron approximation at the scale of $\lambda_{\mathrm{B}}$.
The scale $\lambda_{\mathrm{B}}/a= r_{\mathrm{g}}/\gamma$ corresponds to the synchrotron peak frequency.
On the smallest scale down to $2\pi/k_{\mathrm{max}}$, the trajectory is approximately straight, 
and jittering is responsible for the power law component in the highest frequency region in Figure \ref{spece}.
 }
\label{sasie}
\end{figure}

Next we interpret the spectral features for $a=5$ and 7.
The conceptual diagram of these spectra for $5\leq a<\gamma$ is depicted in Figure \ref{spece}, 
and a schematic picture of an electron trajectory is depicted in Figure \ref{sasie}. 
We explain the appearance of another break in the low frequency range 
 seen at around $10\omega_{\mathrm{g}}$ in Figure \ref{a5}.
On the scale smaller than $\lambda_{\mathrm{B}}$, the electron motion may be approximated by a helical orbit, 
while it is regarded as a randomly fluctuating trajectory when seen on scales larger than $\lambda_{\mathrm{B}}$.
Therefore, for the former scale, we can apply the synchrotron approximation to the emitted radiation.
The beaming cone corresponding to the frequency $\omega$ is given by
\begin{equation}
\theta_{\mathrm{cone}} = \frac{1}{\gamma}(\frac{3\omega_{\mathrm{syn}}}{\omega})^{1/3} 
\label{rlcn}
\end{equation}
(Jackson 1999).
The deflection angle $\theta_0$ of the electron orbit during a time $\lambda_{\mathrm{B}}/c$ is estimated
to be $\theta_0 = a/\gamma$ from the condition
\begin{equation}
\frac{\gamma\lambda_{\mathrm{B}}}{a}\theta_0 = \lambda_{\mathrm{B}}
\end{equation} 
as seen in Figure \ref{sasie}.
Thus, the synchrotron theory is applicable only for $\theta_{\mathrm{cone}} < \theta_0$, so that
the break frequency is determined by $\theta_0 = \theta_{\mathrm{cone}}$, and we obtain
\begin{equation}
\omega_{\mathrm{br}} \sim a^{-3}\omega_{\mathrm{syn}}. 
\label{br}
\end{equation}
This break frequency is the same as obtained by Medvedev (Medvedev 2010).
We understand that as $a$ is larger, break frequency becomes lower, and when $a$ is comparable to $ \gamma$, $\omega_{\mathrm{br}}$ 
coincides with the fundamental frequency $e\sigma/\gamma mc$.

\begin{figure}
\epsscale{.80}
\includegraphics[angle=270]{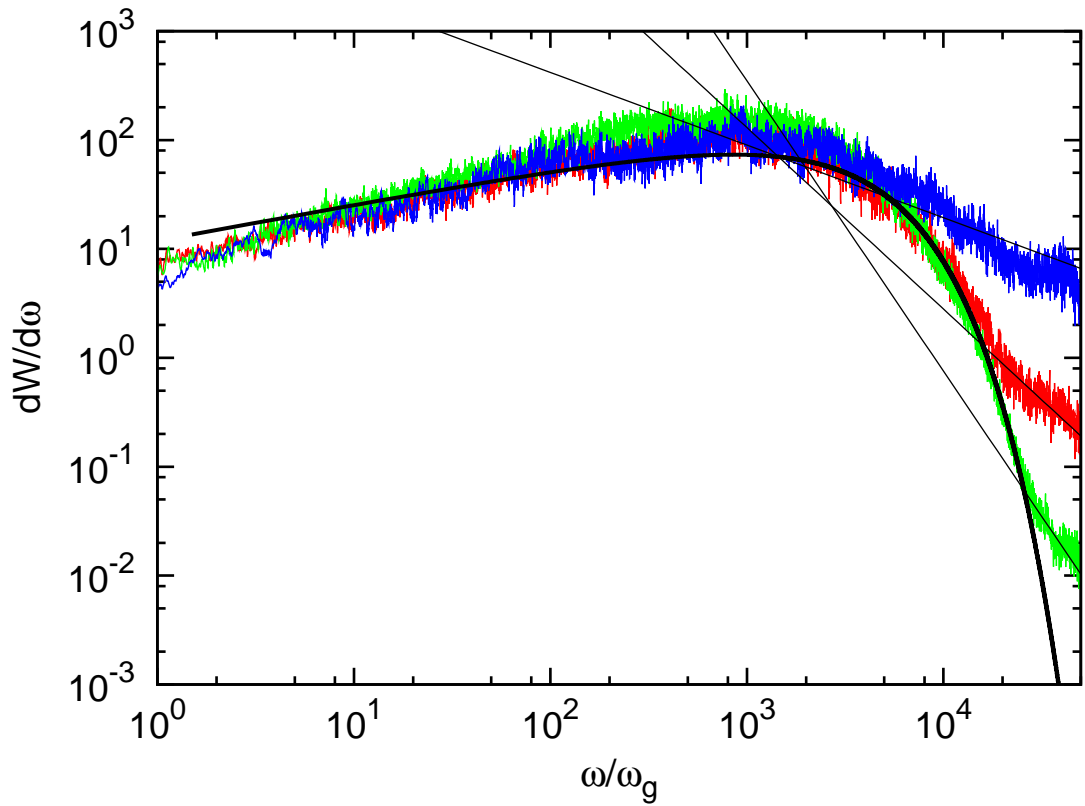}
\caption{Radiation spectra for $a=7\:\mathrm{and}\:\gamma=10$, with $ \alpha = 14/3,11/3\:\mathrm{and}\:8/3$. 
Green, red and blue lines are calculated spectra $\alpha = 14/3,11/3\:\mathrm{and}\: 8/3$, respectively.
Straight black lines are $dW/d\omega \propto \omega^{- \alpha+2}$.
Curved black line is a theoretical curve of synchrotron radiation.
We see effects of different spectrum of the turbulent magnetic fields in high energy region.}
\label{a7}
\end{figure}

Next, we discuss on the high frequency radiation, which results from the electron
 trajectory on scales smaller than $\lambda_{B}$.
The synchrotron theory applies between $\lambda_{\mathrm{B}}/a=r_{\mathrm{g}}/\gamma$
 and $\lambda_{\mathrm{B}}$.
However, we should notice that electron motion suffers from acceleration
 by magnetic turbulence on scales smaller than $\lambda_{\mathrm{B}}/a$.
The trajectory down to the smallest scale of $2\pi/k_{\mathrm{max}}$ is jittering, 
which is attributed to higher wavenumber modes as seen in the zoom up of Figure \ref{sasie}.
If the field in this regime is relatively weak, i.e., $\alpha$ is relatively large (Figure \ref{a7}, green line: $\alpha = 14/3$),
the trajectory on the scale smaller than $\lambda_{\mathrm{B}}/a$ does not much deviate from a helical orbit. 
In this case, radiation spectrum reveals an exponential cutoff, and a power law component appears only in the highest frequency region.
On the contrary, if the smaller scale field is relatively strong, as in the case of $\alpha = 8/3$ depicted in the blue line in Figure \ref{a7},  
the power law component becomes predominant in the high frequency region, and the synchrotron exponential cutoff is smeared out.
The intersection frequency of curved black line and straight black lines at around $10^3\omega_{\mathrm{g}}$ in Figure \ref{a7}
corresponds to $\omega_{\mathrm{jit}}$ as seen in Figure \ref{spece}.
Since the intersection frequency is determined by $a$, the frequency where the power law component appears over the synchrotron cutoff is dependent on $\alpha$.
The excess from the theoretical curve in the middle frequency region in Figure \ref{a7}
may be explained by consideration of two effects.
One is the contribution of hidden DSR component, and the other is a range of synchrotron peak frequency 
which is caused by a fluctuation of magnetic field intensity.  

Fleishman reported that the spectrum for $1<a<\gamma$ and $3<\alpha<4$ becomes a broken power law (Fleishman \& Urtiev 2010).
Medvedev asserted that the high frequency region of the spectrum reveals an exponential cutoff 
for $1<a<\gamma$ (Medvedev 2010).
Our result indicates that an exponential cutoff plus an extra power law component appears, 
which is different from Fleishman's remark and from Medvedev's remark on the high frequency region.
On the other hand, similar spectra to ours have been reported in Fleishman (2005) and 
Reville \& Kirk (2010) when a uniform field is added to turbulent field.
Because the high energy power law component arises from a turbulent spectrum over the wavenumber space,
 this component does not exist when the small scale field 
is excited only in a narrow range of wavenumber space.
Since the energy cascade of turbulent magnetic fields should exist at least to some degree, 
we regard that the higher wavenumber modes naturally exist.
It depends on the set of parameters of $\sigma \: \mathrm{and} \: k_{\mathrm{max}}$ 
whether this high energy power law component can be seen or not.
If $ 2\pi e\sigma/mc^2k_{\mathrm{max}} > 1 $, this component will not be seen.
If the magnetic turbulence is excited by Weibel instability at the relativistic shocks, 
it is not possible for $k_{\mathrm{\max}}$ to be much larger than $k_{\mathrm{\min}}$ 
because the wavelength of injection ($\lambda_{\mathrm{B}} = 2\pi/k_{\mathrm{min}}$) is only a few
 ten times the skin depth at most.
Therefore, the component will not be seen for $a\gg1$ while for $a\sim O(1)$,
this power law component will be seen.

As for the frequency region lower than the break frequency  $\omega_{\rm{br}} = a^{-3} \omega_{\rm{syn}}$,
Medvedev remarked that the spectrum is similar to small angle jitter radiation (Medvedev 2010).
However, it remains to be open if it is so for $1<a<\gamma$, because the assumption that 
the straight orbit approximation of radiating particle is broken.
To predict the exact radiation spectrum of the frequency region lower than the break frequency,
it is necessary to pursue the particle orbit to follow the long term diffusion which is a formidable task.

\section{Summary}
We calculate the radiation spectrum from relativistic
 electrons moving in the small scale turbulent magnetic fields by using the first principle calculation
utilizing the Lienard-Wiechert potential.
We concentrate our calculation on a range of the strength parameter of $1<a<\gamma$. 
We confirm that the spectrum for $a\sim3$ is a broken power law with an index of low energy side 
$\sim 0.5$, and that some GRBs with low energy spectral index harder than synchrotron 
theory predicts may be explained.
Furthermore, we find that the spectrum for $a\sim 7$ takes a novel shape described by a superposition of 
a broken power law spectrum and a synchrotron one.
Especially, an extra power law component appears beyond the synchrotron cutoff in the high frequency region
reflecting magnetic field fluctuation spectrum.
This is in contrast with previous works (Fleishman \& Urtiev 2010, Medvedev 2010). 
Our spectra for $a=5$ and $a=7$ are different from both of them.
We have given a physical reason for this spectral feature.
This novel spectral shape may be seen in various other scenes in astrophysics. 
For example, the spectrum of 3C273 jet at the knot region may be due to this feature (Uchiyama et al. 2006).

We thank the referee for helpful comments. 
We are grateful to T. Okada, S. Tanaka, M. Yamaguchi for discussion and suggestions.
This work is partially supported by KAKENHI 20540231 (F.T.).

\clearpage

\end{document}